\pdfoutput=1
\documentclass{JINST}

\usepackage{graphicx}
\usepackage{dcolumn}
\usepackage{amsmath}

\title{
    Nonparametric Regression using the Concept of Minimum Energy
}

\author{Mike Williams\\
Physics Department, Imperial College London, London, SW7 2AZ, United Kingdom
}

\abstract{
  It has recently been shown that an unbinned distance-based statistic, the {\em energy}, can be used to construct an extremely powerful nonparametric multivariate two sample goodness-of-fit test.  An extension to this method that makes it possible to perform nonparametric regression using multiple multivariate data sets is presented in this paper.  The technique, which is based on the concept of minimizing the energy of the system, permits determination of parameters of interest without the need for parametric expressions of the parent distributions of the data sets.  The application and performance of this new method is discussed in the context of some simple example analyses.  
}

\begin{document}

\section{Introduction}
\label{sec:intro}

The concept of {\em nuissance parameters}, unknown parameters whose values are of no interest but must be determined so that estimators for the parameters of interest can be obtained, is a well known one in high energy physics.  The name is apt in that the presence of such parameters increases the uncertainty on the parameters of interest but has little affect on how the analysis is performed.  
{\em E.g.}, if the functional form of the probability density fuction (p.d.f.) is known, then the unkown parameters in the p.d.f.\ are typically determined using the least squares or maximum likelihood methods.  The presence of nuissance parameters simply increases the number of parameters whose values must be determined but does not affect how the values are obtained.

The same cannot be said for {\em nuissance distributions}, {\em i.e.}, distributions whose functional form is unknown and is of no interest but (seemingly) must be determined to obtain estimators for the parameters of interest.  
The most common solution to this problem in high energy physics is to 
obtain 
an estimate for such a p.d.f.\ either by fitting a model to the data or by binning the data to obtain the integral of the p.d.f.\ inside the bin.   Both of these methods have their limitations: the (often unknown) systematic uncertainties in the model must be propagated back to the parameters, while binning the data results in information loss that tends to increase the statistical uncertainties.  
Another method that is used (albeit less frequently) in high energy physics is nonparametric kernel regression (see, {\em e.g.}, Ref.~\cite{ref:kernel}).  This approach provides an unbinned data-driven way of obtaining an estimate for an unknown p.d.f.; however, its power and reliability is strongly tied to the quality of the p.d.f.\ estimate (which can be difficult to assess).  

All of these methods share the same basic underlying idea: one must obtain an estimate of the p.d.f.\ to obtain estimators for the parameters of interest.  In this paper I will show that this is not the case.  If one has obtained multiple (possibly multivariate) data sets whose p.d.f.'s are known to be related by some set of parameters, then the values of those parameters can be determined without the need for any estimates of the p.d.f.'s themselves; all that is required is the data obtained from the p.d.f.'s.  This paper is laid out as follows: the method is presented in Section~\ref{sec:method}; some example applications are given in Section~\ref{sec:ex} while a summary and discussion is provided in Section~\ref{sec:sum}.

\section{Method}
\label{sec:method}

Consider the case where $n_d$ data sets with completely unknown p.d.f.'s, denoted by $f_1(\vec{x})\ldots f_{n_d}(\vec{x})$, have been obtained (p.d.f.'s are normailzed such that $\int f(\vec{x})d\vec{x} = 1$).  Furthermore, data has also been taken that is known to have a p.d.f.\ that can be written as follows:
\begin{equation}
  \label{eq:f-def}
  f(\vec{x}) = \sum\limits_i^{n_d} \beta_i f_i(\vec{x}), \qquad \sum\limits_i^{n_d} \beta_i = 1,
\end{equation}
where $\beta_i$ are the unknown parameters of interest in the analysis.  The scenario studied in this paper is that the analyst seeks to measure the parameters $\beta_i$ but has no interest in the p.d.f.'s $f_i$.

The following test statistic correlates the difference between two p.d.f.'s at different points in a multivariate space~\cite{ref:baringhaus,ref:aslan}:
\begin{eqnarray}
  \label{eq:energy-def}
  T &=& \frac{1}{2}\int\int \left(f(\vec{x}) - f_0(\vec{x}) \right)  \left(f(\vec{x}^{\prime}) - f_0(\vec{x}^{\prime}) \right) \psi(|\vec{x} - \vec{x}^{\prime}|) d\vec{x} d\vec{x}^{\prime} \nonumber \\
  &=& \frac{1}{2}\int\int \left[f(\vec{x})f(\vec{x}^{\prime}) + f_0(\vec{x})f_0(\vec{x}^{\prime}) -2f(\vec{x})f_0(\vec{x}^{\prime})\right] \psi(|\vec{x}-\vec{x}^{\prime}|) d\vec{x}d\vec{x}^{\prime},
\end{eqnarray}
where $\psi(|\vec{x} - \vec{x}^{\prime}|)$ is a weighting function.  $T$ can be estimated without the need for any knowledge about the forms of $f$ and $f_0$ using data sampled from the p.d.f.'s as
\begin{equation}
  \label{eq:t-calc}
  T \approx \frac{1}{n(n-1)}\sum\limits_{i,j>i}^{n} \psi(\Delta\vec{x}_{ij}) + \frac{1}{n_0(n_0-1)}\sum\limits_{i,j>i}^{n_0} \psi(\Delta\vec{x}_{ij}) 
  - \frac{1}{n n_0}\sum\limits_{i,j}^{n,n_0} \psi(\Delta\vec{x}_{ij}),
\end{equation}
where $\Delta\vec{x}_{ij} =|\vec{x}_i - \vec{x}_j|$ and $n$ ($n_0$) is the number of events sampled from $f$ ($f_0$).  In the order in which they appear in Eq.~\ref{eq:t-calc}, the sums are over pairs of $f$ events, pairs of $f_0$ events and pairs consisting of an $f$ event and an $f_0$ event, respectively. Eq.~\ref{eq:t-calc} is simply Eq.~\ref{eq:energy-def} rewritten using the fact that $\int f(\vec{x}) d\vec{x} = \int f_0(\vec{x})d\vec{x}=1$, along with the standard Monte Carlo integration approximation. 

It is straightforward to calculate $T$ in this way once a metric is chosen that defines distance in the multivariate space (see Ref.~\cite{ref:gof} for a detailed discussion on metrics; this choice has almost no affect on the results). It is worth noting that the larger the difference is between $f$ and $f_0$ the larger the expectation value of $T$ becomes; thus, $T$ can be used to determine the goodness of fit (g.o.f.) of the data to the hypothesis $f = f_0$ (for a more detailed discussion, see Refs.~\cite{ref:baringhaus,ref:aslan,ref:gof}).  From this point forward I will follow Ref.~\cite{ref:aslan} and refer to this test as the energy test (the name originates from the fact that if $\psi(x) = 1/x$ then Eq.~\ref{eq:energy-def} is the electrostatic energy of two charge distributions of opposite sign; the minimum energy occurs when $f = f_0$).

I will now extend the method of Ref.~\cite{ref:baringhaus,ref:aslan} to allow for regression using multiple data sets sampled from different p.d.f.'s.  Consider the following test p.d.f.:
\begin{equation}
  \label{eq:f0-def}
  f_0(\vec{x}, \vec{\beta}) = \sum\limits_i^{n_d} \beta_i f_i(\vec{x}), \qquad \sum\limits_i^{n_d} \beta_i =1,
\end{equation}
where $\beta_i$ are unknown real parameters.  The $T$-value that compares $f$ and $f_0$ in this case is
\begin{eqnarray}
  \label{eq:t-mult}
  T \approx \frac{1}{n(n-1)}\sum\limits_{i,j>i}^{n} \psi(\Delta\vec{x}_{ij}) 
  + \sum\limits_{k}^{n_d}\frac{\beta_k^2}{n_k(n_k-1)}\sum\limits_{i,j>i}^{n_k} \psi(\Delta\vec{x}_{ij})  \hspace{1.5in}\nonumber \\
  \hspace{1.5in}+ \sum\limits_{k,l > k}^{n_d}\frac{\beta_k \beta_l}{n_k n_l}\sum\limits_{i,j}^{n_k,n_l} \psi(\Delta\vec{x}_{ij}) 
  - \sum\limits_k^{n_d}\frac{\beta_k}{n n_k}\sum\limits_{i,j}^{n,n_k} \psi(\Delta\vec{x}_{ij}),
\end{eqnarray}
where $n_k$ is the number of events in the data set sampled from $f_k(\vec{x})$. In the order in which they appear in Eq.~\ref{eq:t-mult} the sums are over pairs of $f$ events, pairs of $f_k$ events, pairs consisting of an $f_k$ event and an $f_l$ event and pairs consisting of an $f$ and an $f_k$ event, respectively.
The values of $\beta_k$ that minimize $T$, $\hat{\beta}_k$, provide the best g.o.f.  In the limit $n,n_k \to \infty$ Eq.~\ref{eq:t-mult} approaches Eq.~\ref{eq:energy-def} and $T = 0$ if $\hat{\beta}_k = \beta_k$  for each $k$.  It is important to note that Eq.~\ref{eq:f0-def} is defined using properly normalized p.d.f.'s.  If unnormalized functions are used, then the estimators are related to the true values by $\hat{\beta}_k = \beta_k \int f_k(\vec{x})d\vec{x}$.
The uncertainty on the fit parameter values is easily obtained using a resampling technique ({\em e.g.}, bootstrapping).


One might think that calculating $T$ takes so much CPU time that running this fit is not practical; however, a careful inspection of Eq.~\ref{eq:t-mult} reveals that the CPU-intensive components of $T$, the $\sum \psi(\Delta\vec{x})$ terms, do not depend on the $\beta_k$ values.  Thus, these terms only need to be calculated once (which can be done prior to running the fit).  
A limitation of this method, however, is that if a p.d.f.\ contributes negative probability to $f_0$, then the fits may be unstable due to fact that it is not possible to constrain $f_0$ to be non-negative for all $\vec{x}$.  In many cases it will be easy to tell if this is a problem or not.  Otherwise, I recommend performing a Monte Carlo study prior to using this method in such situations. 
{\em N.b.}, in Section~\ref{sec:multi-ex} an example is presented where $n_d = 3$ and one p.d.f.\ contributes negative probability to $f_0$ and the method still works properly (in that example, the known relationships between the p.d.f.'s makes it easy to determine that the p.d.f.\ is non-negative everywhere even though one term contributes negative probability). 

In the next section I will present two examples to help illustrate how the method works.  The main goal will be to not only demonstrate how to use the method but to also solidify in the reader's mind what the $\hat{\beta}_k$ values represent.  I conclude this section by restating the main result of this work: estimators for the parameters $\beta_k$ in Eq.~\ref{eq:f-def} can be obtained by minimizing Eq.~\ref{eq:t-mult} using data sets sampled from $f_k$ without the need for any knowledge about the p.d.f.'s themselves.

\section{Example Applications}
\label{sec:ex}
The examples I will present in this section are meant to demonstrate how the method works.  From a physics perspective, these are not the most interesting applications of the method; however, they are effective at illustrating how to use the method and how to interpret the meaning of the fit parameters.  I will first present a very simple univariate example followed by an example using Dalitz plots.  
The weighting function that will be used for the examples below is $\psi(x) = -\log{(x + \epsilon)}$, where $\epsilon$ is of the order of the inverse of the total number of events in all samples combined ($\epsilon$ simply guards against an infinite contribution due to two extremely close events; the exact value is not important).  See Section~\ref{sec:sum} for discussion on the choice of weighting function.

\subsection{Univariate Example}
\label{sec:uni-ex}

Consider the case where one has three data sets sampled from the p.d.f.'s $f_c(x) = 1$, $f_l(x) = 2x$ and $f(x) = (ax+b)/(a/2+b)$ on the interval $x \in [0,1)$.  The sample sizes of these data sets are denoted by $n_c$, $n_l$ and $n$, respectively.  The p.d.f.\ $f$ can be rewritten in terms of the other two p.d.f.'s as 
\begin{equation}
\label{eq:simple-pdf}
f(x) = \frac{\beta_1 f_l(x) + \beta_0 f_c(x)}{\beta_1 + \beta_0},
\end{equation}
where $\beta_1 = a/2$, $\beta_0 = b$ and the normalization requirement on the $\beta$ values has been made explicit.  The values $\beta_1/(\beta_1 + \beta_0)$ and $\beta_0/(\beta_1 + \beta_0)$ represent the fraction of $f$'s probability associated with $f_l$ and $f_c$, respectively.  {\em E.g.}, if $f_l$ and $f_c$ represent signal and background p.d.f.'s then the signal purity is given by $\beta_1/(\beta_1 + \beta_0)$.  The factor of 2 in $\beta_1$ arises because the function $x$ is not a properly normalized p.d.f.\ on this interval.  

The fit procedure is simple: first, all of the $\sum \psi(\Delta x_{ij})$ terms in Eq.~\ref{eq:t-mult} need to be calculated.  This can be time consuming for large data sets but it only needs to be done once.  The test p.d.f.\ is then written as 
\begin{equation}
  f_0(x,\vec{\beta}) = \frac{\beta_1 f_l(x) + \beta_0 f_c(x)}{\beta_1 + \beta_0},
\end{equation}
where no knowledge about the forms of $f_l$ and $f_c$ is assumed to be known and the normalization requirement on the $\beta$ values has again been made explicit.  Calculating $T$ using Eq.~\ref{eq:t-mult} is then straightforward for any given values of $\beta_1$ and $\beta_0$.  The estimators for $\beta_i$ are the values that minimize $T$, $\hat{\beta}_i$.  

\begin{figure}
  \centering
  \includegraphics[width=0.4\textwidth]{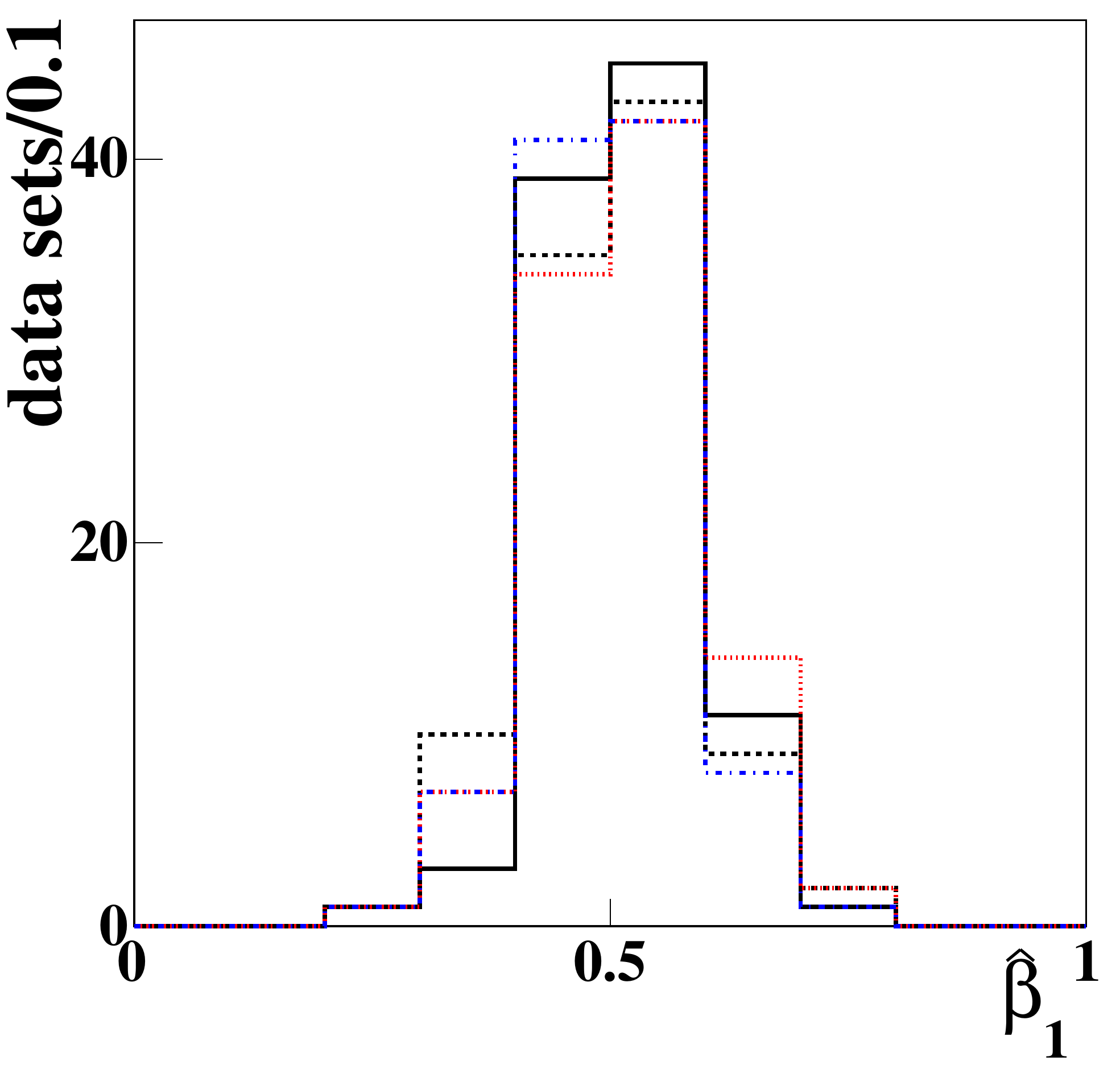}
  \caption[]{\label{fig:simple}
    $\hat{\beta}_1$ values obtained for the ensemble of data sets described in Section~\ref{sec:uni-ex} using the energy test for $n_{c,l}=1$k (black, dashed) and $n_{c,l}=10$k (black, solid).  Results obtained using the binned $\chi^2$ test (red, dotted) and maximum likelihood test (blue, dash-dotted) are also shown.  
  }
\end{figure}

I generated an ensemble of 100 data sets from Eq.~\ref{eq:simple-pdf} using the values $\beta_0 = \beta_1 = 1/2$ (or, equivalently, $a=1, b=1/2$).  I chose $n=500$ and two different values for $n_c$ and $n_l$: $n_{c,l}=1$k and $n_{c,l}=10$k. Figure~\ref{fig:simple} shows the  $\hat{\beta}_1$ values ($\hat{\beta}_0$ is just $1-\hat{\beta}_1$) obtained for these data sets using the minimum energy, $\chi^2$ and maximum likelihood tests (statistics are given in Tab.~\ref{tab:simple}).  The $\chi^2$ and maximum likelihood tests were performed by fitting the function $f_0(x) \propto \beta_1 x+\beta_0$ to the data sets sampled from $f$; {\em i.e.}, the data sets sampled from $f_c$ and $f_l$ were not used and, instead, the functional form of $f$ was assumed to be known.  Even though I {\em cheated} ({\em i.e.}, used information I assumed I do not actually have access to) using these tests, the performance of the energy test is still comparable.  The method works: I have determined the fraction of probability in $f$ from $f_c$ and $f_l$ without having any knowledge of the forms of $f_c$ or $f_l$ and without trying to approximate $f_c$ and $f_l$ using the data ({\em e.g.}, using a kernel-based method).

\begin{table}
  \begin{center}
  \begin{tabular}{|cc|cc|cc|cc|}
    \hline
    \multicolumn{2}{|c|}{$\chi^2$ test} &  \multicolumn{2}{c|}{$\log{\mathcal{L}}$ test} & \multicolumn{4}{c|}{energy test}\\
    \multicolumn{2}{|c|}{} & \multicolumn{2}{|c|}{} & \multicolumn{2}{c|}{$n_{c,l}=1$k} &  \multicolumn{2}{c|}{$n_{c,l}=10$k}\\
    $\mu$ & r.m.s. & $\mu$ & r.m.s. & $\mu$ & r.m.s. & $\mu$ & r.m.s.\\
    \hline
    0.52 & 0.09 & 0.50 & 0.08 & 0.50 & 0.09 & 0.51 & 0.08 \\
    \hline
  \end{tabular}
  \caption[]{\label{tab:simple}
    $\hat{\beta}_1$ statistics obtained for the ensemble of data sets described in Section~\ref{sec:uni-ex} using the $\chi^2$, maximum likelihood and energy tests.
    The data sets were sampled from a p.d.f.\ with a value of 0.5.
  }
  \end{center}
\end{table}

\begin{figure}
  \centering
  \includegraphics[width=0.4\textwidth]{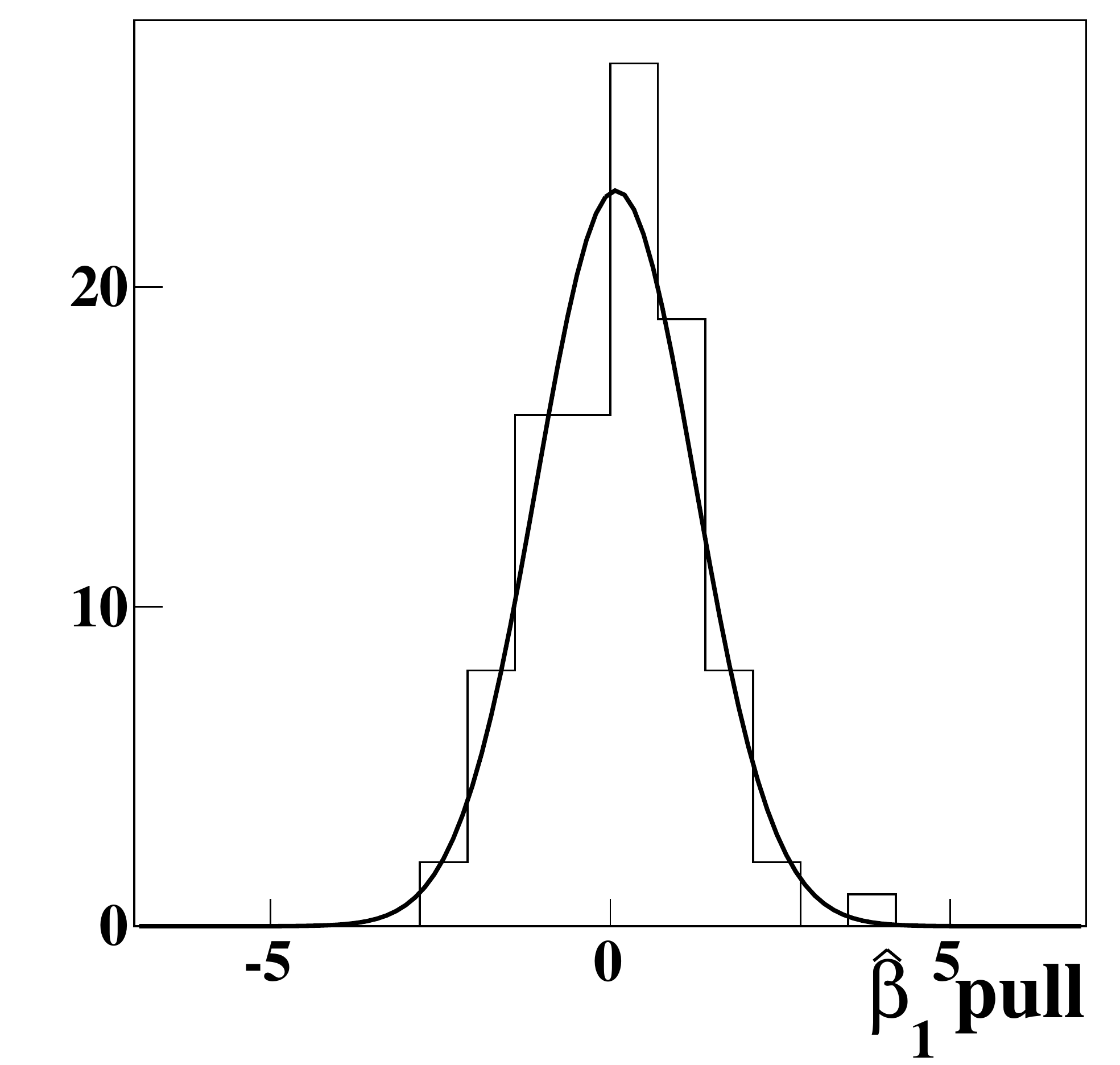}
  \caption[]{\label{fig:simple-pull}
    $\hat{\beta}_1$ pull distribution obtained for the $n_{c,l}=10$k data sets where the parameter uncertainties are estimated using the bootstrapping technique.  The solid line shows a fit to a gaussian distribution; the fit yields $\mu = 0.08 \pm 0.13$ and $\sigma = 1.16 \pm 0.11$.
  }
\end{figure}

As stated above, the uncertainty on the fit parameters must be obtained using a data resampling technique.  There are many such techniques available;  I chose to use bootstrapping.  This technique involves making $n_{\rm boot}$ (I chose 100) resampled data sets from each $f$, $f_c$ and $f_l$ data set.  These {\em bootstrap copy} data sets are produced by sampling with replacement (thus, the sample sizes are unchanged) from the originals.  The fit parameter $\hat{\beta}_1$ is then computed for each of the $n_{\rm boot}$ bootstrap data sets; the standard deviation of this distribution is used as an estimate for the uncertainty on $\hat{\beta}_1$.  Figure~\ref{fig:simple-pull} shows the $\hat{\beta}_1$ pull distribution obtained for the ensemble of data sets described in the previous paragraph.  The uncertainty on $\hat{\beta}_1$ was determined for each data set in the ensemble using bootstrapping.  The pull distribution is consistent with the expected standard normal distribution; thus, I conclude that the bootstrap method produces reliable estimates for the uncertainty on the $\hat{\beta}_k$ values.  For more information on bootstrapping, see Ref.~\cite{ref:efron}.  


\subsection{Multivariate Example}
\label{sec:multi-ex}

I will now consider a multivariate example involving Dalitz plots.  The decay $D \to K_S \pi^+ \pi^-$ is allowed for $D = D^0,\; \bar{D}^0$ and any superposition of $D^0$ and $\bar{D}^0$.  Experimentally, flavor-tagged samples ({\em i.e.}, pure $D^0$ or $\bar{D}^0$) can be obtained using the decays $D^{*+} \to D^0 \pi^+$ and $D^{*-} \to \bar{D}^0 \pi^-$ since the charge of the slow pion tags the flavor of the $D$. Such data sets with sample sizes of $\mathcal{O}(100{\rm k})$ have been obtained at Belle~\cite{ref:belle}.  $CP$ eigenstates can be obtained using quantum correlated $D^0 \bar{D}^0$ pairs produced in the reaction $e^+ e^- \to D^0 \bar{D}^0$.  $CP$-tagged data sets have been obtained at CLEO with sample sizes of several hundreds of events~\cite{ref:cleo}, while sample sizes of $\mathcal{O}(1000)$ have been recorded (but not yet published) at BES~\cite{ref:bes}.  

\begin{figure}
  \centering
  \includegraphics[width=0.4\textwidth]{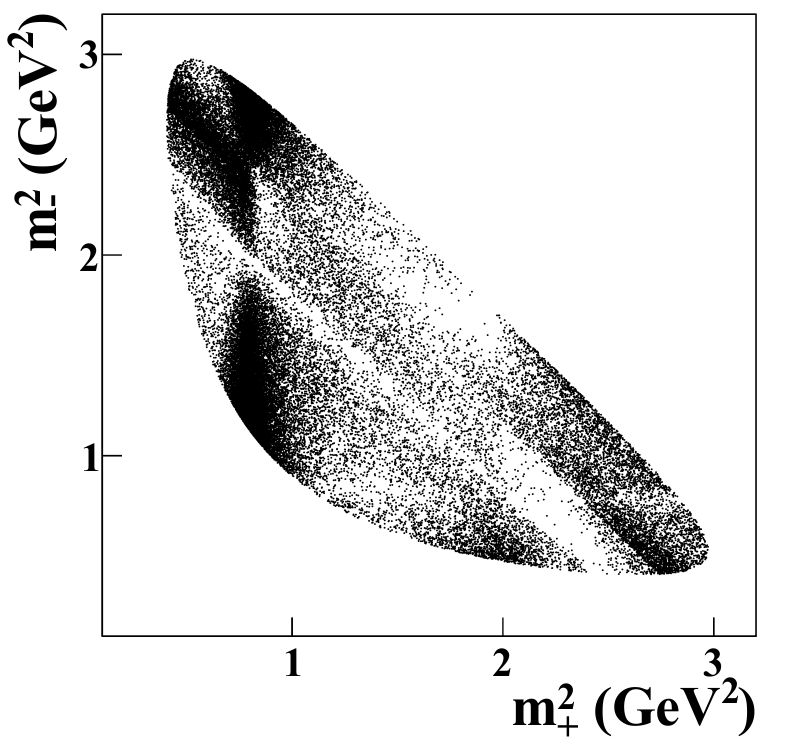}
  \includegraphics[width=0.4\textwidth]{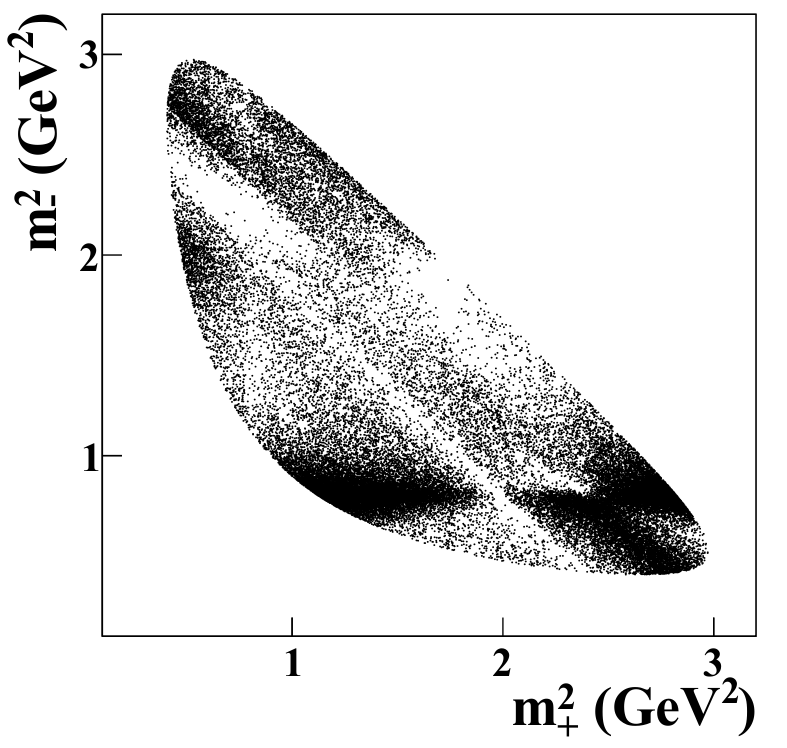} \\
  \includegraphics[width=0.4\textwidth]{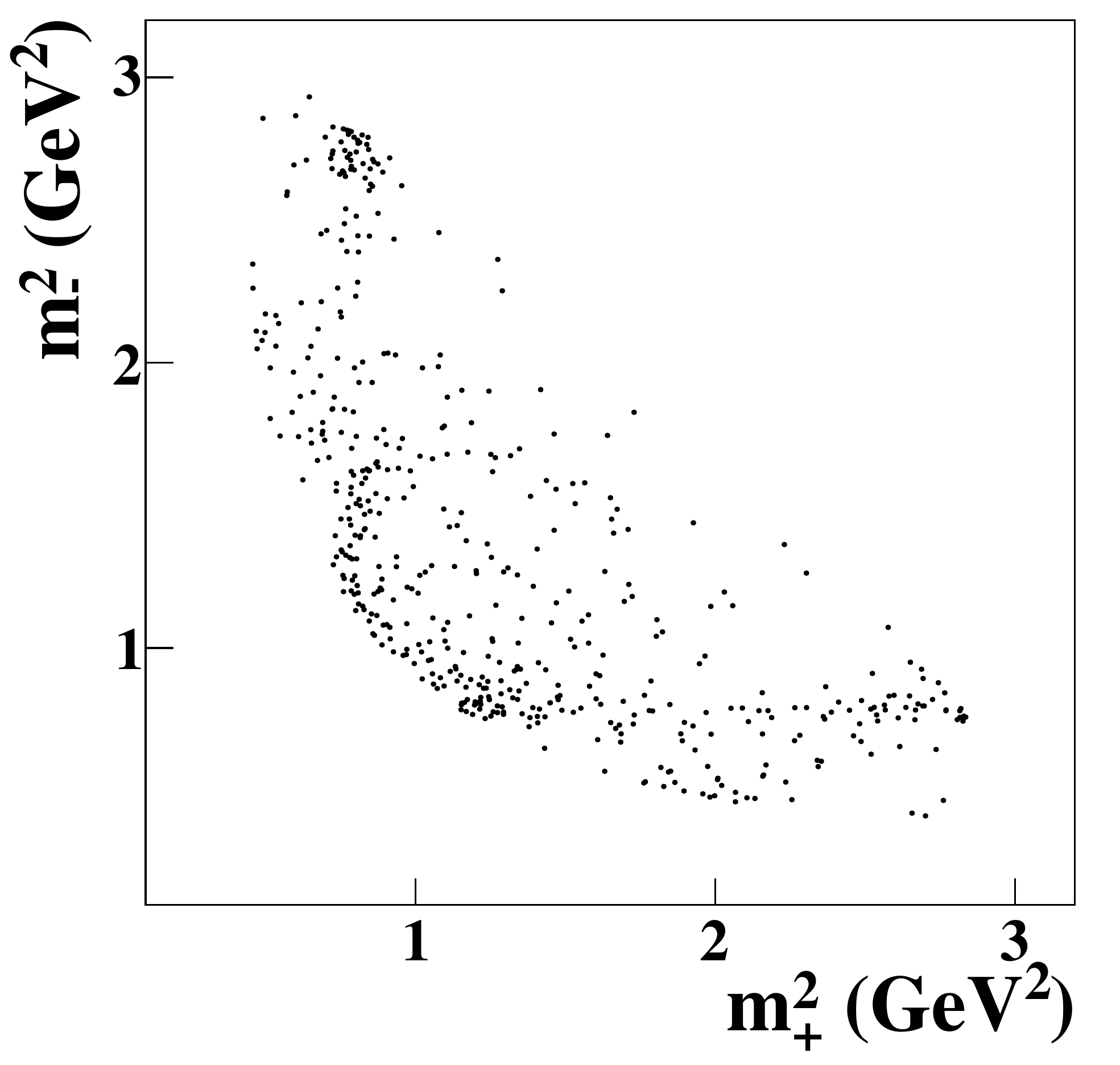}
  \includegraphics[width=0.4\textwidth]{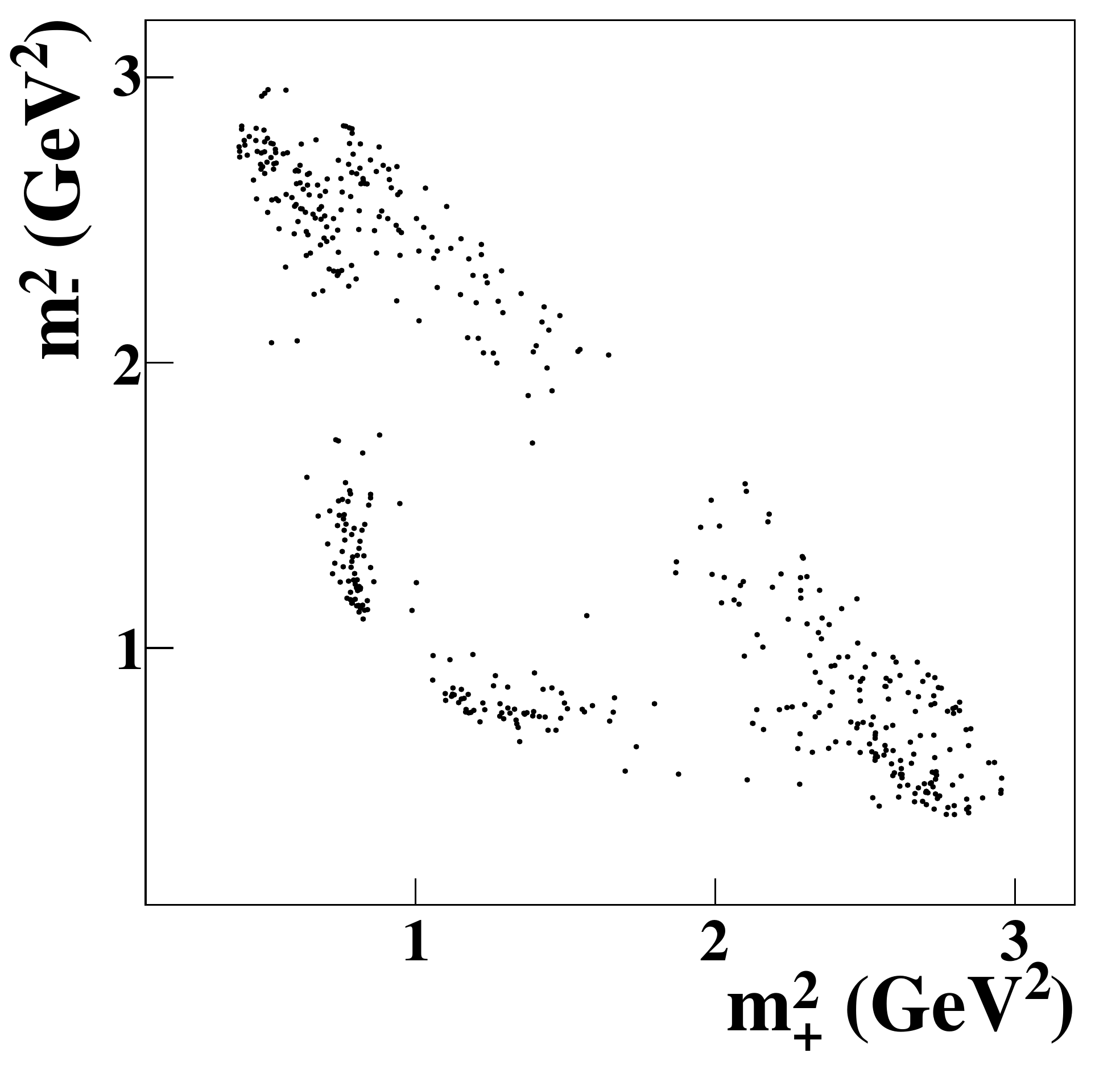}
  \caption[]{\label{fig:dps}
    Dalitz plots for $D \to K_S \pi^+ \pi^-$ for (top left) $\bar{D}^0$, (top right) $D^0$, (bottom left) $D_{CP+}$ and (bottom right) $D_{CP-}$.
  }
\end{figure}

I will proceed under the assumption of no $CP$ violation in the neutral $D$ system, which is known to be valid to a high level of precision.  I will denote the quantum mechanical amplitudes for the decays $D^0, \bar{D}^0 \to K_S \pi^+ \pi^-$ as follows:
\begin{eqnarray}
  \mathcal{A}_{\bar{D}^0 \to K_S \pi^+ \pi^-}(\vec{x}) &\equiv& \bar{\mathcal{A}}(\vec{x}), \\
  \mathcal{A}_{D^0 \to K_S \pi^+ \pi^-}(\vec{x}) &\equiv& \mathcal{A}(\vec{x}),
\end{eqnarray}
where $\vec{x} = (m_+^2,m_-^2)$ and $m_+^2$ and $m_-^2$ are the invariant masses of the $K_S \pi^+$ and $K_S \pi^-$ systems, respectively.  The flavor-tagged p.d.f.'s are then given by
\begin{eqnarray}
  \bar{f}(\vec{x}) = |\bar{\mathcal{A}}(\vec{x})|^2/\mathcal{I},  \\
  f(\vec{x}) = |\mathcal{A}(\vec{x})|^2/\mathcal{I},
\end{eqnarray}
where $\mathcal{I} = \int |\bar{\mathcal{A}}(\vec{x})|^2 d\vec{x} = \int |\mathcal{A}(\vec{x})|^2 d\vec{x}$ (which is valid in the absence of $CP$ violation).  
The amplitudes for the case where the $D$ is tagged to be in a $CP$ eigenstate are given by
\begin{equation}
  \mathcal{A}_{\pm}(\vec{x}) = \frac{1}{\sqrt{2}}[\bar{\mathcal{A}}(\vec{x}) \pm \mathcal{A}(\vec{x})].
\end{equation}
The $CP$-tagged p.d.f.'s are thus
\begin{equation}
  \label{eq:cp-pdf}
  f_{\pm}(\vec{x}) = \left(|\bar{\mathcal{A}}(\vec{x})|^2 + |\mathcal{A}(\vec{x})|^2 \pm 2 |\bar{\mathcal{A}}(\vec{x})||\mathcal{A}(\vec{x})|\cos \Delta\theta(\vec{x})\right)/\mathcal{I}_{\pm},
\end{equation}
where 
\begin{equation} 
\mathcal{I}_{\pm} = 2\left(\mathcal{I} \pm \int |\bar{\mathcal{A}}(\vec{x})||\mathcal{A}(\vec{x})|\cos \Delta\theta(\vec{x}) d\vec{x}\right)
\end{equation} 
and $\Delta\theta(\vec{x})$ is the phase difference between $\bar{\mathcal{A}}(\vec{x})$ and $\mathcal{A}(\vec{x})$ at each point $\vec{x}$.
I generated an ensemble of 100 data sets for each flavor- and $CP$-tagged decay using the model of Ref.~\cite{ref:belle} (the amplitudes were evaluated using the {\tt qft++} package~\cite{ref:qft++}).  Figure~\ref{fig:dps} shows an example of a data set of each tagging type.  The sample sizes were chosen to be $n = \bar{n} = 100$k and $n_{\pm} = 1000$ for the $D^0$, $\bar{D}^0$ and $D_{CP\pm}$ decays, respectively.  

As an example of using the method presented in this paper, consider the case where one has collected both flavor-tagged and $CP$-tagged data sets.  I will assume that the only knowledge about $f(\vec{x})$ and $\bar{f}(\vec{x})$ is the fact that  $\int |\bar{\mathcal{A}}(\vec{x})|^2 d\vec{x} = \int |\mathcal{A}(\vec{x})|^2 d\vec{x}$; the functional forms of $f$ and $\bar{f}$ are completely unknown.  I will also assume that the $CP$-odd tagged data set is known to follow Eq.~\ref{eq:cp-pdf}; however, only the relative coefficients of $|\bar{\mathcal{A}}(\vec{x})|^2$, $|\mathcal{A}(\vec{x})|^2$ and $|\bar{\mathcal{A}}(\vec{x})||\mathcal{A}(\vec{x})|\cos \Delta\theta(\vec{x})$ are known (nothing is known about their functional forms).  For the $CP$-even data set, I assume that its p.d.f. can be written as
\begin{equation}
  \label{eq:cp-f}
  f_+(\vec{x}, \vec{\alpha}) \propto \alpha_0 |\bar{\mathcal{A}}(\vec{x})|^2 + \alpha_1 |\mathcal{A}(\vec{x})|^2 + \alpha_2 |\bar{\mathcal{A}}(\vec{x})||\mathcal{A}(\vec{x})|\cos \Delta\theta(\vec{x}).
\end{equation}
The goal of the analysis is to test the quality of the $CP$ tagging by determining the $\alpha_i$ values (up to an arbitrary normalization factor) using only the available knowledge about the data collected, which does not include any knowledge about the functional forms of $\bar{\mathcal{A}}(\vec{x})$ or $\mathcal{A}(\vec{x})$.  A deviation from the $CP$-even values, $\vec{\alpha} \propto (1,1,2)$, would signify a problem in the $CP$ tagging.

The energy test can be used to determine the $\beta_i$ values in the following test p.d.f. using the same procedure described in the previous section:
\begin{equation}
  \label{eq:cp-f0}
  f_{0}(\vec{x}, \vec{\beta}) = \beta_0 \bar{f}(\vec{x}) + \beta_1 f(\vec{x}) + \beta_2 f_-(\vec{x}), \qquad \sum \beta_i = 1,
\end{equation}
where $f$, $\bar{f}$ and $f_-$ are the flavor-tagged and $CP$-odd-tagged p.d.f.'s, respectively (as defined above).
Comparing Eq.~\ref{eq:cp-f0} and Eq.~\ref{eq:cp-f} one can see that the $\vec{\beta}$ values are related to the $\vec{\alpha}$ values by
\begin{equation}
  \label{eq:a2b}
  \vec{\alpha} \propto \left(\beta_0 + \beta_2 \mathcal{I}/\mathcal{I}_-,\beta_1 + \beta_2 \mathcal{I}/\mathcal{I}_-,  - 2 \beta_2 \mathcal{I}/\mathcal{I}_-\right).
  \end{equation}
Thus, to determine the coefficients of $|\bar{\mathcal{A}}(\vec{x})|^2$, $|\mathcal{A}(\vec{x})|^2$ and $|\bar{\mathcal{A}}(\vec{x})||\mathcal{A}(\vec{x})|\cos \Delta\theta(\vec{x})$ for the $CP$-even data set ({\em i.e.}, to determine $f_+(\vec{x})$ using the energy test) requires the value of $\mathcal{I_-}/\mathcal{I}$ to be known (or obtainable).  At first glance one might think that this is a showstopper; however, the ratio of these integrals can be obtained from the sample sizes obtained for the various data sets.  

Ref.~\cite{ref:cleo} measured both the flavor- and $CP$-tagged yields using quantum-correlated $D^0\bar{D}^0$ pairs; thus,  $\mathcal{I_-}/\mathcal{I}$ can be obtained using
\begin{equation}
  \label{eq:int-ratio}
  \frac{\mathcal{I}}{\mathcal{I}_-} = \frac{(n^{\prime}/n_D + \bar{n}^{\prime}/n_{\bar{D}})/2}{2 n_-/n_{CP-}},
\end{equation}
where $n_D$, $n_{\bar{D}}$ and $n_{CP-}$ are the total number of $D^0$, $\bar{D}^0$ and  $CP$-odd tagged quantum-correlated $D^0\bar{D}^0$ pairs, respectively, and $n^{\prime}, \bar{n}^{\prime}$ are the number of flavor-tagged $D^0,\bar{D}^0 \to K_S \pi \pi$ decays observed at CLEO (not the ones observed at Belle;  the Belle data are used in the fits because they have larger sample sizes).  
The sample sizes are $n_D \sim 100 n^{\prime}$, $n_{\bar{D}} \sim 100 \bar{n}^{\prime}$ and $n_{CP-} \sim 100 n_-$. The uncertainties on the various yield ratios in Eq.~\ref{eq:int-ratio} are binomial; thus, the statistical uncertainty on $\mathcal{I}/\mathcal{I}_-$ is negligible due to the fact that the total number of $CP$-tagged quantum-correlated $D^0\bar{D}^0$ pairs is 100$\times$ larger than the $CP$-tagged $D \to K_S \pi \pi$ samples. 

\begin{figure}
  \centering
  \includegraphics[width=0.4\textwidth]{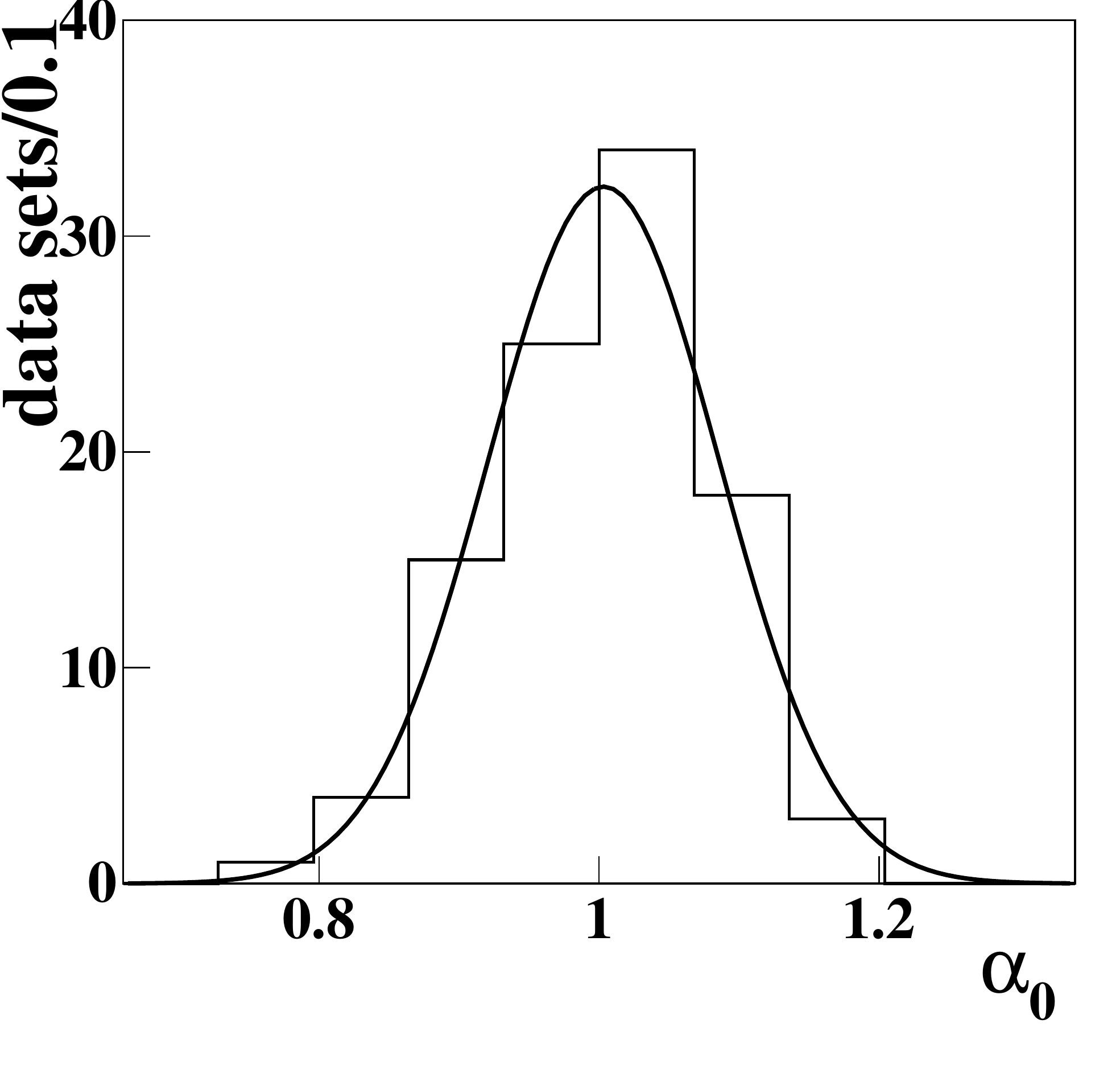}
  \includegraphics[width=0.4\textwidth]{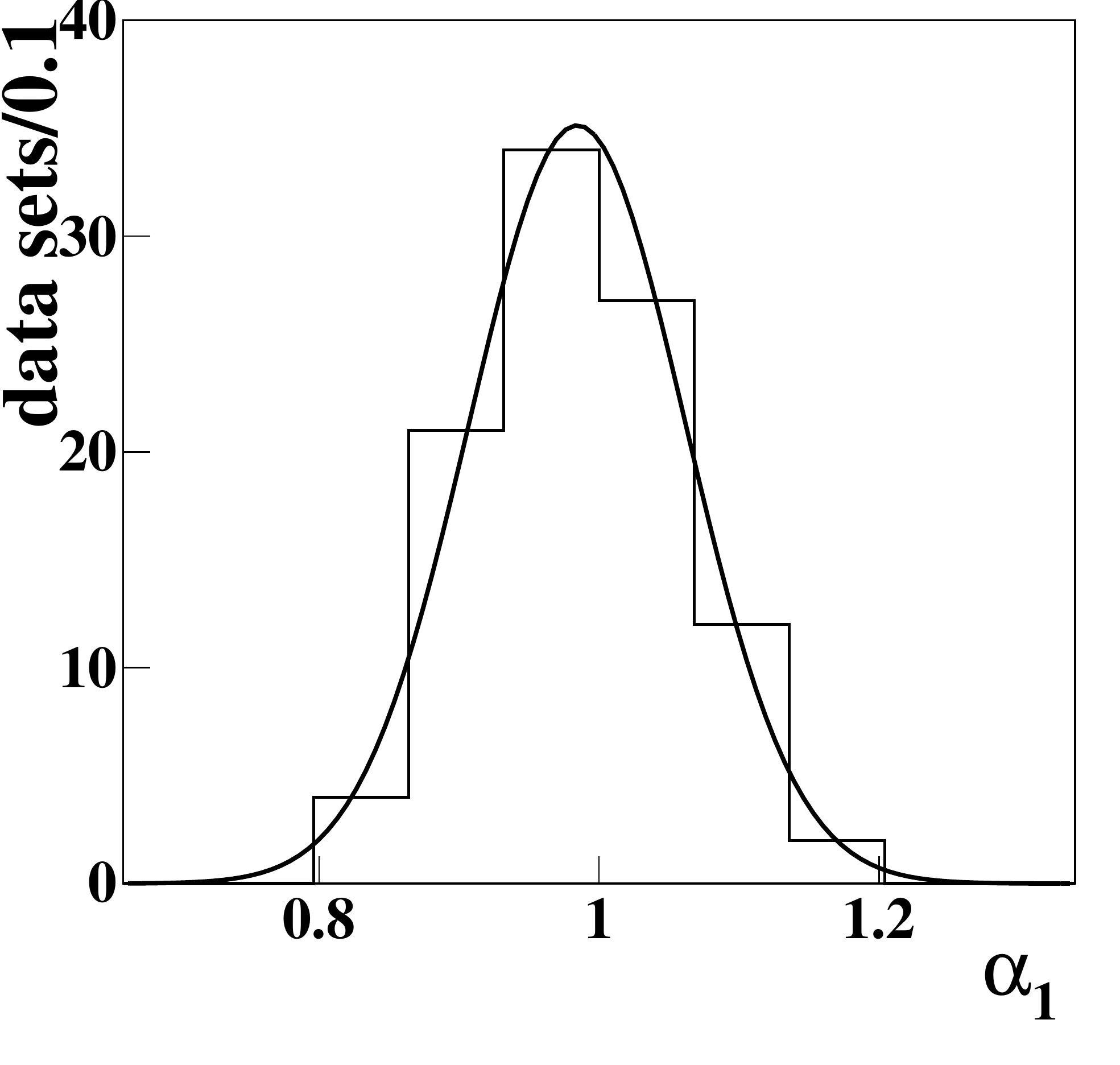} \\
  \includegraphics[width=0.4\textwidth]{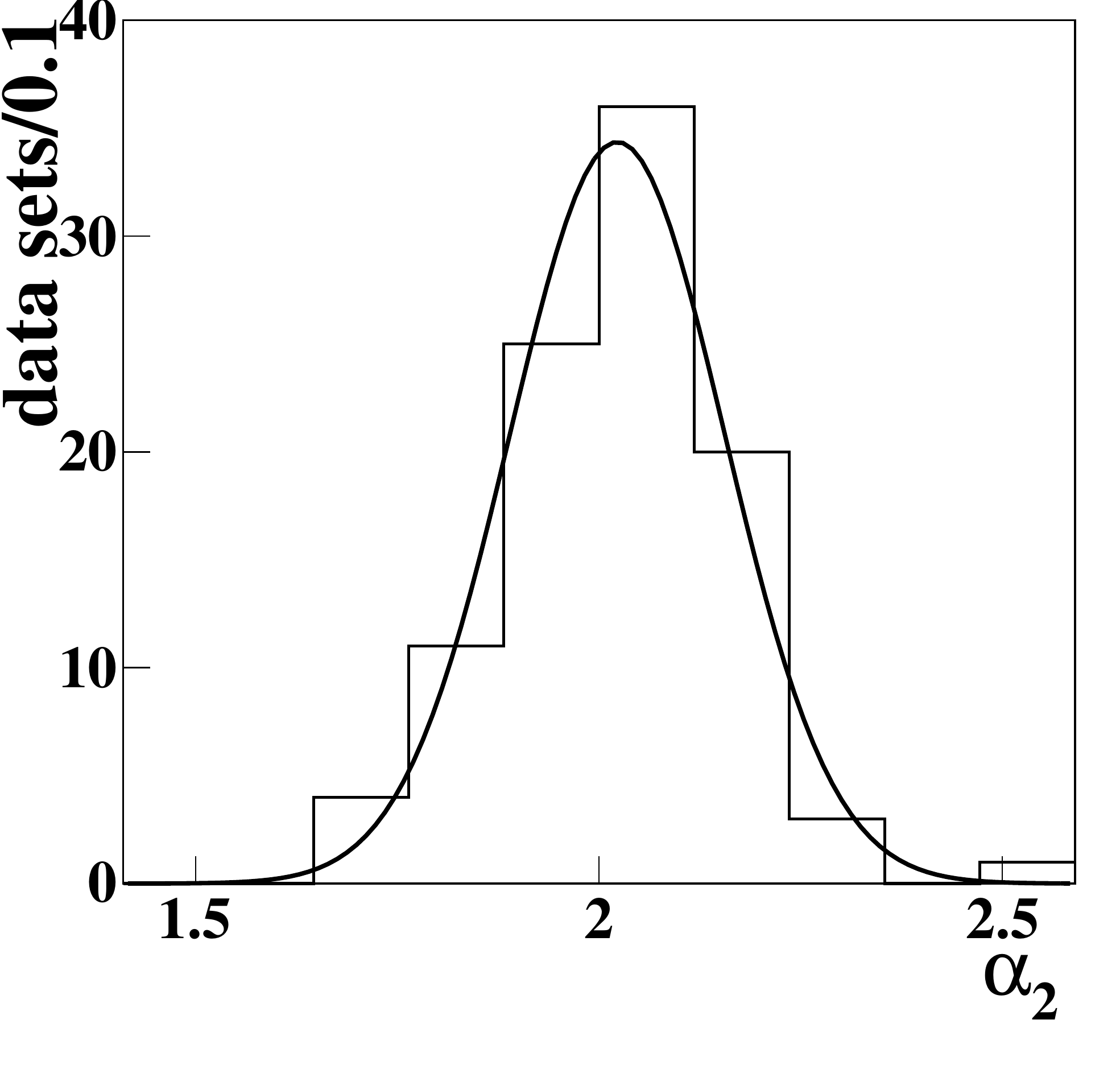}
  \caption[]{\label{fig:alpha}
    Values obtained for $\vec{\alpha}$ in Eq.~\ref{eq:cp-f} using the energy test.  The solid lines shows fits of gaussian distributions to the data.  The resulting means and widths are: $\mu_0 = 1.00 \pm 0.01$, $\sigma_0 = 0.07 \pm 0.01$; $\mu_1 = 0.98 \pm 0.01$, $\sigma_1 = 0.07 \pm 0.01$ and $\mu_2 = 2.02 \pm 0.02$, $\sigma_2 = 0.14 \pm 0.02$.
  }
\end{figure}

Figure~\ref{fig:alpha} shows the results obtained using the energy test to determine $\vec{\beta}$ and Eq.~\ref{eq:a2b} to convert these into $\vec{\alpha}$ values.  The results are in excellent agreement with the true values $\vec{\alpha} = (1,1,2)$.  The relative statistical uncertainty obtained on the $\alpha$ values is 7\% (see Section~\ref{sec:sum} for discussion on improving this resolution).  Looking at Fig.~\ref{fig:dps} one can imagine that obtaining a model p.d.f.\ for this analysis with a small systematic uncertainty requires a lot of work.  Since the goal of this example was to simply obtain estimators for $\vec{\alpha}$ (to test the quality of the $CP$ tagging), there is no reason to build such a model.  Instead, one can just use the energy test.

\section{Background \& Efficiency}

In these examples I have ignored two very important experimental issues: (1) the existence of background events and (2) the presence of non-uniform detector efficiency effects.   Both of these can be accounted for in this method by introducing a weighting factor, $w$, for each event.  The weight factor is simply given by $w_i = P^i_S/P^i_D$, where $P_S^i$ and $P_D^i$ are the probabilities that the event is signal (and not background) and is detected, respectively.   Eq.~\ref{eq:t-mult} would then need to be rewritten as follows:
\begin{eqnarray}
  \label{eq:t-mult-bkgd}
  T \approx \frac{1}{W^2}\sum\limits_{i,j>i}^{n} w_i w_j \psi(\Delta\vec{x}_{ij}) 
  + \sum\limits_{k}^{n_d}\frac{\beta_k^2}{W_k^2}\sum\limits_{i,j>i}^{n_k} w_i w_j \psi(\Delta\vec{x}_{ij})  \hspace{1.5in}\nonumber \\
  \hspace{1.5in}+ \sum\limits_{k,l > k}^{n_d}\frac{\beta_k \beta_l}{W_k W_l}\sum\limits_{i,j}^{n_k,n_l} w_i w_j \psi(\Delta\vec{x}_{ij}) 
  - \sum\limits_k^{n_d}\frac{\beta_k}{W W_k}\sum\limits_{i,j}^{n,n_k} w_i w_j\psi(\Delta\vec{x}_{ij}),
\end{eqnarray}
where $W$ and $W_k$ are the sum of the weight factors for the $f$ and $f_k$ data sets, respectively.  If $w_i = 1$ for all events in all data sets then Eq.~\ref{eq:t-mult} is recovered (up to a negligible change, $n(n-1) \to n^2$, in the same data set sums; this change has been made solely for ease of notation and could easily be omitted if desired).   
{\em N.b.}, in many cases these weights can be left out.  
It is up to the analyst to decide when these factors are important and to include them when necessary.

\section{Summary \& Discussion}
\label{sec:sum}
In this paper I have shown that the concept of minimum energy can be used to carry out nonparametric regression.  The parameter values that are obtained give the fractional probability associated with each p.d.f.  For many analyses, {\em e.g.}, signal-background subtraction, this is sufficient.  To convert these to coefficients of unnormalized functions requires inputting the ratio of the integrals of these functions.  These ratios can often times be obtained from measured event yields (an example of which was given in Section~\ref{sec:multi-ex}).  I have also shown that data resampling methods, {\em e.g.}, bootstrapping, can be used to obtain good estimates of the uncertainties on the parameter values. 

It is well known that in many binned analyses specialized binning schemes can be developed to enhance the resolution on the parameters of interest of the $\chi^2$ test.  The same is true for the energy test with regards to the weighting function $\psi(x)$.  In this paper I chose a simple generic $\psi(x)$ because the goal was to demonstrate how the method works; however, for many analyses it will be possible to obtain a better resolution by choosing a $\psi(x)$ tailored to the problem being analyzed.  Refs.~\cite{ref:baringhaus,ref:aslan} discuss several possible choices for $\psi(x)$ and there are undoubtedly many other possibilities as well; however, for most analyses (where the p.d.f.'s vary in a smooth and slow way) the choice used in this paper is likely to be close to optimal (whatever that may be).

\section*{Acknowledgements}
I would like to thank Ulrik Egede, Tim Gershon and Vladimir Gligorov for discussions.
This work is supported by the Science and Technology Facilities Council (United Kingdom) under grant number ST/H000992/1.

\end{document}